\documentclass[preprint,showpacs,preprintnumbers,amsmath,amssymb,showkeys]{revtex4}

\usepackage[unicode,bookmarks,bookmarksopen,bookmarksopenlevel=0,colorlinks,%
pageanchor=1,breaklinks=1,linkcolor=blue,citecolor=blue,urlcolor=red,hypertex]{hyperref}

\usepackage{graphicx}
\usepackage{dcolumn}
\usepackage{bm}

\usepackage{float}

\begin{document}

\preprint{APS/123-QED}

\title{Mathematical Model of a pH-gradient Creation at~Isoelectrofocusing. \\ Part IV: Theory}


\author{E.\,V.~Shiryaeva}
\email{shir@math.sfedu.ru}

\author{N.\,M.~Zhukova}
\email{zhuk_nata@mail.ru}

\author{M.\,Yu.~Zhukov}%
 \email{myuzhukov@gmail.com}
\affiliation{%
Southern Federal University\\ Rostov-on-Don, Russia
}%

\date{\today}

\begin{abstract}
The mathematical model describing the non-stationary natural $\textrm{pH}$-gradient arising under the action of an electric field in an aqueous solution of ampholytes (amino acids)  is constructed. The model is a part of a more general model of the isoelectrofocusing (IEF) process. The presented model takes into account: 1) general Ohm's law (electric current flux includes the diffusive electric current); 2) dissociation of water; 3) difference between isoelectric point (IEP) and isoionic  point (PZC -- point of zero charge). We also study the Kohlraush's function evolution and discuss the role of the Poisson-Boltzmann equation.

\end{abstract}


\maketitle

\section{Introduction}\label{ZhS-01}

This paper continues the series of papers \cite{Part1,Part2,Part3} about $\textrm{pH}$-gradient creation at isoelectrofocusing (IEF). Here, the general mathematical model of IEF is obtained.
The construction of accurate electrophoresis mathematical models is described in the works \cite{BabZhukYudE,ZhukovBabskiyYudovich,MosherSavilleThorman,Thormann2004,Thormann2006,Zhukov2005},
in which, in particular, are described and classified the various methods of electrophoresis: zone electrophoresis, isotachophoresis, and isielectrofocusing. The models presented in these works are either very general or, on the contrary, describe very partial problems. Usually, at constructing isoelectrofocusing model the simplifying assumptions are chosen.  In particular, dissociation of water is not always taken into account, Ohm's law does not include terms that corresponds  to diffusion current, \emph{etc}. All such simplification can lead to the violation of the basic physical laws such as the law of conservation of mass or the law of conservation of electric charge. Despite the fact that the differences between isoelectric point (IEP) and isoionic (PZC) point are already described in \cite{BabZhukYudE,ZhukovBabskiyYudovich}, this effect is usually omitted at the constructing IEF. In \cite{Part3}  the results of the numerical investigation for IEF model are presented. In this work it is shown that  differences between IEP and PZC
take the important role, especially for almost stationary regime. Mentioned effects are considered in this paper. Of course, the model is not complete. In particular, the influence of the ionic strength of a solution on the mobility of ions, the effect of Wien and others are not taken into account.

The paper is organized as follows. In Sec.~\ref{App:ZhS-6} we demonstrate the method of the general model constructing.
In Sec.~\ref{App:ZhS-7.1n} we study the role of the difference between the isoelectric and isoionic points.
In Sec.~\ref{App:ZhS-7.1} we obtain and study the Kohlraush's function.
Finally, in Sec.~\ref{App:ZhS-7.2} we discuss the role of the Poisson-Boltzmann equation.



\section{Mathematical model of IEF}\label{App:ZhS-6}

To construct the mathematical model of IEF we use the theory of the local chemical equilibrium described in \cite{ZhukovBabskiyYudovich,BabZhukYudE,Zhukov2005}.
Generally, it is convenient to write the dissociation reactions for a solution consisting of $n$ amphoteric substances as ($k=1,\dots,n$)
\begin{equation}\label{ZhSeq-F1}
   a_k^{+}   \overset{B_k}{\rightleftharpoons}   a_k^{0} + \textrm{H}^{+},   \quad
   a_k^{0}   \overset{A_k}{\rightleftharpoons}   a_k^{-} + \textrm{H}^{+},
\end{equation}
where $a_{k}^{0}$ is a zwitterion (`neutral' ion), $A_k$ and $B_k$ are dissociation constants for acid
group $a_{k}^{-}$ (negative ion) and based group $a_{k}^{+}$ (positive ion) correspondingly.

The chemical kinetic equations have the following form
\begin{equation}\label{ZhSeq-F2}
     \frac{da_k^{+}}{dt}=r_k^{+},   \quad \frac{da_k^{-}}{dt}=r_k^{-}, \quad \frac{da_k^{0}}{dt}=r_k^{0},
\end{equation}
where
\begin{equation}\label{ZhSeq-F3}
     r_k^{+}= - B_k^{+}  a_k^{+}  + B_k^{-} a_k^{0} [\textrm{H}^{+}],
\end{equation}
\begin{equation*}
     r_k^{-}= - A_k^{+}  a_k^{0}  + A_k^{-} a_k^{-} [\textrm{H}^{+}],
\end{equation*}
\begin{equation*}
     r_k^{0}= -r_k^{+}-r_k^{+}.
\end{equation*}
Here, $a_k^{+}$, $a_k^{-}$, $a_k^{0}$ are the molar concentration; $r_k^{+}$, $r_k^{-}$, $r_k^{0}$ are the density of concentration  sources (we use the same symbol for denotation of the substance and its concentration), $[\textrm{H}^{+}]$ is the analytical concentration of the hydrogen ion; $A_k^{+}$, $B_k^{+}$, $A_k^{-}$  $A_k^{-}$ are the velocities of the direct and reverse reactions.

The ion concentrations are connect to analytical concentration of the substance $a_k$ with the help of the relations:
\begin{equation}\label{ZhSeq-F4}
     a_k^{+}= \theta_k^{+} a_k,   \quad
     a_k^{-}= \theta_k^{-} a_k,  \quad
     a_k=a_k^{0} + a_k^{+} + a_k^{-},
\end{equation}
where $\theta_k^{-}$, $\theta_k^{+}$ are the dissociation degrees (further, we show that the dissociation degrees are depend on $[\textrm{H}^{+}]$ only, \emph{i.\,e.}
$\theta_k^{+}=\theta_k^{+}([\textrm{H}^{+}])$, $\theta_k^{-}=\theta_k^{-}([\textrm{H}^{+}]))$.

The mass transport under action of an electric field is described by the equations (dimensionless variables):
\begin{equation}\label{ZhSeq-F5}
\partial_t a_{k}^{+}+\operatorname{div} \boldsymbol{i}_k^{+}=r_{k}^{+},\quad \boldsymbol{i}_k^{+} = -\varepsilon\mu_{k}^{+} \nabla a_k^{+}+z_{k}^{+}\mu_{k}^{+} \theta_k^{+}([\textrm{H}^{+}])a_k^{+} \boldsymbol{E},
\quad k=1,\dots,n,
\end{equation}
\begin{equation*}
\partial_t a_{k}^{-}+\operatorname{div} \boldsymbol{i}_k^{-}=r_{k}^{-},\quad \boldsymbol{i}_k^{-} = -\varepsilon\mu_{k}^{-} \nabla a_k^{-}+z_{k}^{-}\mu_{k}^{-} \theta_k^{-}([\textrm{H}^{+}])a_k^{-} \boldsymbol{E},
\quad k=1,\dots,n,
\end{equation*}
\begin{equation*}
\partial_t a_{k}^{0}+\operatorname{div} \boldsymbol{i}_k^{0}=r_{k}^{0},\quad \boldsymbol{i}_k^{0} = -\varepsilon\mu_{k}^{0} \nabla a_k^{0},
\quad k=1,\dots,n,
\end{equation*}
where $\boldsymbol{i}_k^{-}$, $\boldsymbol{i}_k^{+}$, $\boldsymbol{i}_k^{0}$ are the flux densities, $\boldsymbol{E}$ is the intensity of electric field,
$z_{k}^{-}\mu_{k}^{-}$, $z_{k}^{+}\mu_{k}^{+}$, $\varepsilon\mu_{k}^{-}$, $\varepsilon\mu_{k}^{+}$ are the electrophoretic mobilities and diffusive coefficients of the ions,
$\varepsilon\mu_{k}^{0}$ is the diffusive coefficients of the `neutral' ion, $z_{k}^{-}=-1$, $z_{k}^{+}=+1$ are the ion charges (ion charge per unit of the electron charge).

We assume that
\begin{equation}\label{ZhSeq-F6}
\mu_{k}^{-}=\mu_{k}^{+}=\mu_k, \quad
\varepsilon\mu_{k}^{0}=\varepsilon\mu_{k}^{-}=\varepsilon\mu_{k}^{+}=\varepsilon \mu_k,
\quad k=1,\dots,n.
\end{equation}

In this case the summation equations (\ref{ZhSeq-F5}) for each $k$ give (see (\ref{ZhSeq-F3}), (\ref{ZhSeq-F4}))
\begin{equation}\label{ZhSeq-F7}
\partial_t a_{k}+\operatorname{div} \boldsymbol{i}_k=0,
\quad k=1,\dots,n,
\end{equation}
\begin{equation}\label{ZhSeq-F8}
\boldsymbol{i}_k = -\varepsilon\mu_{k} \nabla a_k+\mu_{k} \theta_{k} a_k \boldsymbol{E},
\quad k=1,\dots,n,
\end{equation}
where
\begin{equation}\label{ZhSeq-F9}
     a_k=a_k^{0} + a_k^{+} + a_k^{-},
\end{equation}
\begin{equation*}
    \boldsymbol{i}_k=\boldsymbol{i}_k^{-}+\boldsymbol{i}_k^{0}+\boldsymbol{i}_k^{+},
\end{equation*}
\begin{equation*}
     \theta_k=\theta_k([\textrm{H}^{+}])=\theta_k^{+}-\theta_k^{-}.
\end{equation*}

Here, $\theta_k$ is the specific molar charge of the substance $a_k$.

Note, the equations (\ref{ZhSeq-F7}) do not contain the density of source. Other words, these equations are the conservative laws (not balance equations). The analytical concentrations $a_k$ are integrals of the chemical kinetic equations (\ref{ZhSeq-F2}).

The system of equations (\ref{ZhSeq-F7}), (\ref{ZhSeq-F8}) is apparent, however we add two comments on the related physical
processes. First, the equations are written for the concentrations $a_k$. It means that these equations describe the
distributions of some complex chemical substances consisting of ions and zwitterions (not ions and
neutral substances separately). From the physical viewpoint we deal only with such substances (not with their
components) and we can not (without employing of special methods) observe the components of the $k$-th substance. Second, the
system (\ref{ZhSeq-F7}), (\ref{ZhSeq-F8}) is so called unclosed system, because the molar charge $\theta_k$ is undefined (even if temporarily assume that the intensity of the electric field $\boldsymbol{E}$ is given).

To close the system (\ref{ZhSeq-F7}), (\ref{ZhSeq-F8}) we use hypothesis of the local chemical equilibrium introduced in \cite{ZhukovBabskiyYudovich,BabZhukYudE} and developed in \cite{Zhukov2005}. We assume that dissociation chemical reactions are very fast (which are completed almost instantly). It allows to believe that the  conditions of the chemical equilibrium are valid:
\begin{equation}\label{ZhSeq-F10}
      r_k^{-}=0, \quad r_k^{+}=0, \quad (r_k^{0}=0),
\end{equation}
or
\begin{equation}\label{ZhSeq-F11}
     \frac{a_k^{0} [\textrm{H}^{+}]}{a_k^{+}}=\frac{B_k^{+}}{B_k^{-}}=B_k,   \quad
     \frac{a_k^{-} [\textrm{H}^{+}]}{a_k^{0}}=\frac{A_k^{+}}{A_k^{-}}=A_k.
\end{equation}
Solving this system we get the dependance of the dissociation degrees on hydrogen ion concentration
\begin{equation}\label{ZhSeq-F12}
     \theta_k^{+}([\textrm{H}^{+}])=\frac{[\textrm{H}^{+}]^2}{[\textrm{H}^{+}]^2+B_k [\textrm{H}^{+}] + A_k B_k},   \quad
     \theta_k^{-}([\textrm{H}^{+}])=\frac{A_k B_k}{[\textrm{H}^{+}]^2+B_k [\textrm{H}^{+}] + A_k B_k}.
\end{equation}
We emphasize once more that this relations appear as the result of extremely fast chemical reactions that are instant (mathematically) or much faster
than any transfer processes (physically). In fact, we have two types of variables: fast variables ($a_k^{0}$, $a_k^{-}$, $a_k^{+}$) and slow variables ($a_k$).
The equations (\ref{ZhSeq-F4}), (\ref{ZhSeq-F9})--(\ref{ZhSeq-F11}) give connections  between these variables.

For further we need more additional equations. We must obtain the equation for determining the electric field and the concentration of hydrogen ions.

In the general case in addition to the reactions (\ref{ZhSeq-F1}) one should take into account the
autodissociation of water (as well as the autoionization of water or autoprotolysis)
\begin{equation}\label{ZhSeq-F13}
   \textrm{H}_2\textrm{O} {\,\overset{k^+}{{\underset{k^-}{\rightleftharpoons}}}\,}
     \textrm{H}^{+} + \textrm{OH}^{-}.
\end{equation}
At the local chemical equilibrium the  concentrations of hydrogen ions (or hydronium ions) $[\textrm{H}^+]$ and hydroxide ions  $[\textrm{OH}^-]$ are
reflated as
\begin{equation}\label{ZhSeq-F14}
   [\textrm{OH}^-]=\frac{k^2_w}{[\textrm{H}^+]},
\end{equation}
where $k^2_w$ is the autodissociation constant of water (the synonyms are: ionization constant, dissociation
constant, self-ionization constant, and ion product of water; in dimensional variables
$k_w=10^{-7}\,\textrm{mol/l}$; it should be noticed here that water also represents an amphoteric substance.).

In chemistry the term autodissociation constant is used for $K_w=k_w^2$. However, theoretically it
represents a confusion, since the standard dimension of the dissociation constant is $\textrm{mol/l}$. This is exactly
the dimension of $k_w$ (not $K_w$).

The electroneutrality equation has the following form
\begin{equation}\label{ZhSeq-F15}
   \sum_{k=1}^{n} (a_{k}^{+}-a_{k}^{-})
   +[\textrm{H}^{+}]-[\textrm{OH}^{-}]
   =0.
\end{equation}
Taking into account (\ref{ZhSeq-F4}) we get
\begin{equation}\label{ZhSeq-F16}
   \sum_{k=1}^{n}
   \left\{
   \theta_k^{+}([\textrm{H}^{+}])a_{k}^{+}-\theta_k^{-}([\textrm{H}^{+}])a_{k}^{-}
   \right\}+[\textrm{H}^{+}]-[\textrm{OH}^{-}]=0.
\end{equation}
This equation allows to determine the concentration $[\textrm{H}^{+}]$.

Obviously, the electric current flux densities of the ions have the following form
\begin{equation}\label{ZhSeq-F17}
\boldsymbol{j}_{k}^{+}=z_{k}^{+}\boldsymbol{i}_{k}^{+}=
-\varepsilon \mu_k\nabla a_k^{+}+\mu_{k}^{+}a_{k}^{+} \boldsymbol{E}, \quad
\boldsymbol{j}_{k}^{-}=z_{k}^{+}\boldsymbol{i}_{k}^{+}=
+\varepsilon \mu_k\nabla a_k^{-}-\mu_{k}^{-}a_{k}^{-} \boldsymbol{E},
\end{equation}
\begin{equation}\label{ZhSeq-F18}
\boldsymbol{j}^{H}=
-\varepsilon \mu_{_H}\nabla [\textrm{H}^{+}]+\mu_{_H} [\textrm{H}^{+}] \boldsymbol{E}, \quad
\boldsymbol{j}^{OH}=
+\varepsilon \mu_{_{OH}}\nabla [\textrm{OH}^{-}]-\mu_{_{OH}} [\textrm{OH}^{-}] \boldsymbol{E}, \quad
\end{equation}
where
$\mu_{_H}$, $\mu_{_{OH}}$, $\varepsilon\mu_{_H}$, $\varepsilon\mu_{_{OH}}$ are the electrophoretic mobilities and diffusive coefficients of the water ions.

Then, the electric current flux densities of the mixture is
\begin{equation}\label{ZhSeq-F19}
\boldsymbol{j}=\sum_{k=1}^{n}
\left(
\boldsymbol{j}_{k}^{+}+\boldsymbol{j}_{k}^{-}
\right)+\boldsymbol{j}^{H}+\boldsymbol{j}^{OH}.
\end{equation}

Taking into account (\ref{ZhSeq-F4}) we get
\begin{equation}\label{ZhSeq-F20}
\boldsymbol{j}=\sum_{k=1}^{n}
\left(
-\varepsilon \mu_k\nabla(\theta_{k}a_k)+\mu_{k}\sigma_ka_k \boldsymbol{E}
\right)+
\end{equation}
\begin{equation*}
+
\left(
-\varepsilon \mu_{_H}\nabla [\textrm{H}^{+}]+\mu_{_H} [\textrm{H}^{+}] \boldsymbol{E}
+\varepsilon \mu_{_{OH}}\nabla [\textrm{OH}^{-}]-\mu_{_{OH}} [\textrm{OH}^{-}] \boldsymbol{E}
\right),
\end{equation*}
where
\begin{equation}\label{ZhSeq-F21}
     \theta_k=\theta_k([\textrm{H}^{+}])=\theta_k^{+}-\theta_k^{-}, \quad  \sigma_k=\sigma_k([\textrm{H}^{+}])=\theta_k^{+}+\theta_k^{-}.
\end{equation}
Here, $\theta_k$ is the specific molar charge of the substance $a_k$, $\sigma_k$ is the specific molar conductivity of the substance $a_k$.

The constitutive relation (\ref{ZhSeq-F20} ) is so called generalized Ohm's law which differs from the usual law by the presence of the diffusion terms (see also \cite{Part3}).

The current flux density $\boldsymbol{j}$ satisfy to  	
the equation of the electric current continuity
\begin{equation}\label{ZhSeq-F22}
 \operatorname{div} \boldsymbol{j}=0.
\end{equation}

We also assume that the electric field is potential
\begin{equation}\label{ZhSeq-F23}
  \boldsymbol{E}=-\nabla \varphi,
\end{equation}
where $\varphi$ is the electric potential.

The equations (\ref{ZhSeq-F7}), (\ref{ZhSeq-F8}), (\ref{ZhSeq-F16}), (\ref{ZhSeq-F22}), (\ref{ZhSeq-F23}), and  constitutive relations  (\ref{ZhSeq-F12}), (\ref{ZhSeq-F14}), (\ref{ZhSeq-F20}), (\ref{ZhSeq-F21}) are the complete system of equation that allows to determine the concentrations $a_k$, $[\textrm{H}^{+}]$, $[\textrm{OH}^{-}]$, and potential $\varphi$.

An important characteristic of the solution of amphoteric substances is $\textrm{pH}$; its value is defined by the
concentration of hydrogen ions ${\textrm{H}^+}$ expressed in $\textrm{mol}/\textrm{l}$ with the use of the relation
\begin{equation*}
   \textrm{pH}=-\lg [\textrm{H}^+].
\end{equation*}
It is better to write this expression as
\begin{equation}\label{ZhSeq-F24}
  \textrm{pH}=-\lg \left(\frac{[\textrm{H}^+]}{k_w}\right),
\end{equation}
where $k_w$ is the autodissociation constant of water.

In addition, instead of conventionally used function pH (that represents
the measure of the acidity or alkalinity of a solution) we use acidity function $\psi$ (that is linearly connected with pH),
which is better adapted to our mathematical model
\begin{equation}\label{ZhSeq-F25}
   [\textrm{H}^+]=k_w e^{\psi}, \quad [\textrm{OH}^-]=k_w e^{-\psi},
\end{equation}
\begin{equation*}
   \textrm{pH}=-\lg k_w-\psi\lg e.
\end{equation*}

Usually, the value of $\textrm{pH}$ varies from $0$ to $14$, which corresponds to the changes of $\psi$
in the interval from $-16.118$ to $+16.118$.

In chemistry, instead of dissociation constants $A_k$ and $B_k$ use their negative decimal degrees $\textrm{pA}_i$, $\textrm{pB}_i$ that are given by
relations:
\begin{equation}\label{ZhSeq-F26}
   \textrm{pA}_k=-\lg \left(\frac{A_k}{k_w}\right), \quad \textrm{pB}_k=-\lg \left(\frac{B_k}{k_w}\right).
\end{equation}

We especially emphasize, that the replacement of the concentrations $[\textrm{H}^+]$ and $[\textrm{OH}^-]$ by the acidity function $\psi$ allows to write the system of equations in final form convenient for further mathematical investigation:
\begin{equation}\label{ZhSeq-F27}
\partial_t a_{k}+\operatorname{div} \boldsymbol{i}_k=0,\quad k=1,\dots,n,
\end{equation}
\begin{equation*}
\boldsymbol{i}_k = -\varepsilon\mu_{k} \nabla a_k+\mu_{k} \theta_{k}(\psi)a_k \boldsymbol{E},
\quad k=1,\dots,n,
\end{equation*}
\begin{equation}\label{ZhSeq-F28}
 \sum_{k=1}^{n}\theta_{k}(\psi)a_{k}+2K_w \sinh\psi=0,
\end{equation}
\begin{equation}\label{ZhSeq-F29}
 \operatorname{div} \boldsymbol{j}=0, \quad   \boldsymbol{E}=-\nabla \varphi,
\end{equation}
where
\begin{equation}\label{ZhSeq-F30}
\boldsymbol{j}=\sum_{k=1}^{n}
\left(
-\varepsilon \mu_k\nabla(\theta_{k}(\psi)a_k)+\mu_{k}\sigma_k(\psi)a_k \boldsymbol{E}
\right)+
\end{equation}
\begin{equation*}
+2k_w\mu_0\left(
-\varepsilon
\nabla(\sinh(\psi-\psi_0))+\cosh(\psi-\psi_0) \boldsymbol{E}
\right),
\end{equation*}
\begin{equation}\label{ZhSeq-F31}
\theta_k(\psi)=\frac{\sinh(\psi-\psi_{k})}{\cosh(\psi-\psi_{k})+\delta_{k}}=\frac{\varphi'_k(\psi)}{\varphi_k(\psi)},
\end{equation}
\begin{equation*}
\sigma_k(\psi)=\frac{\cosh(\psi-\psi_{k})}{\cosh(\psi-\psi_{k})+\delta_{k}}=\frac{\varphi''_k(\psi)}{\varphi_k(\psi)},
\end{equation*}
\begin{equation*}
\varphi_k(\psi)=\cosh(\psi-\psi_{k})+\delta_{k},
\end{equation*}
\begin{equation}\label{ZhSeq-F32}
   \psi_k=\frac12 \ln\frac{A_k B_k}{k_w^2}, \quad \delta_k=\frac12 \sqrt{\frac{B_k}{A_k}},
\end{equation}
\begin{equation*}
 \mu_{0}=\sqrt{\mu_{_H} \mu_{_{OH}}}, \quad
 \psi_0=\frac12\ln \frac{\mu_{_{OH}}}{\mu_{_H}},
\end{equation*}
where  $\psi_i$ is the isoelectric point (electrophoretic mobility $\mu_i \theta_i$ is equal to zero at $\psi=\psi_i$, \emph{i.e.} $\mu_i \theta_i(\psi_i)=0$),
$\mu_0$ is the effective mobility of water ions,   $\mu_{_H}$, $\mu_{_{OH}}$ are the mobilities of hydrogen  ${\textrm H}^{+}$ and hydroxide $\textrm{OH}^{-}$ ions,
$\psi_0$ is the value of $\psi$ when water conductivity is minimal, $\delta_i>0$ is the dimensionless parameter, $\varphi_k(\psi)$ is some auxiliary function.

Specify connection parameters $\psi_k$ and $\delta_k$ to the parameters used in chemistry
\begin{equation}\label{ZhSeq-F33}
\textrm{pI}_k=\frac12(\textrm{pA}_k+\textrm{pB}_k), \quad
\textrm{pA}_k-\textrm{pB}_k=2\lg(2\delta_k).
\end{equation}
Here $\textrm{pI}_k$ is the electrophoretic point of amphoteric substance.

The system (\ref{ZhSeq-F27})--(\ref{ZhSeq-F32}) allows to determine concentrations $a_k$, acidity function $\psi$, and electrical potential $\varphi$  (or electric field intensity $\boldsymbol{E}$) when parameters $\varepsilon$, $\mu_k$, $\mu_0$, $\psi_k$, $\psi_0$, $\delta_k$, $k_w$ are given.

We should add comments on the roles of different
equations in the system (\ref{ZhSeq-F27})--(\ref{ZhSeq-F32}) as
well as different terms of these equations. The term
$2k_w \mu_{0} \cosh(\psi-\psi_0)$ in (\ref{ZhSeq-F30}) describes the contribution of water ions into
the mixture conductivity, while the term $2k_w\sinh\psi$ in  (\ref{ZhSeq-F28}) corresponds to the
contribution of these ions into the mixture molar charge.
As a rule, the contribution of water ions to the mixture conductivity and the charge of the mixture is small enough, and these terms for the simplified models can be omitted
(see, for example, \cite{Part1,Part2,Part3}).

The \emph{algebraic} equation (\ref{ZhSeq-F28}) represents the
condition of the electroneutrality of mixture; it allows us
to find $\psi$. In fact, this equation describes the
instant control of medium properties (electrophoretic
mobilities and molar conductivities) by the function $\psi$
(that is linked to the concentration of hydrogen ions or
$\textrm{pH}$ of mixture).

We also assume that the maximal values of
concentrations $a_k$ and the values $\mu_0$, $\psi_0$,
$\mu_k$, $\psi_k$ are all of the order
$O(1)$, while the parameters $\varepsilon$ and
$k_w$ are small. We should also mention on some important properties of the physical processes.
The absence of the concentration flux ($\boldsymbol{i}_k=0$) does not
mean that the $k$-th substance does not participate into an electric current.

For example, let us neglect the diffusion
and take $\varepsilon=0$ in (\ref{ZhSeq-F27}) and (\ref{ZhSeq-F30}). Then at the isoelectric point (when $\psi=\psi_k$) the
charge $\theta_k(\psi_k)a_k=0$ and the mobility $\mu_k\theta_k(\psi_k)=0$, so we get $i_k=0$. At the same time the density
of the electric current at $\psi=\psi_k$ is $\boldsymbol{j}_k=\mu_k\sigma_k(\psi_k)a_k\ne 0$.
However this fact does not contain any contradiction since there are two equal (at $\psi=\psi_k$) but
opposite fluxes $\boldsymbol{i}_k^{-}$ and $\boldsymbol{i}_k^{+}$ of the negative and positive ions that both are driven by the electric field. The
flux of $a_k$ is $\boldsymbol{i}_k=\boldsymbol{i}_k^{+}+\boldsymbol{i}_k^{-}=0$. The density of the electric current in this case is
$\boldsymbol{j}_k=z_k^{+} \boldsymbol{i}_k^{+}+z_k^{-}\boldsymbol{i}_k^{-}\ne 0$. This fact plays a key role in the describing of the transport processes under action of an electric field.

\subsection{Difference between mobilities of the negative and positive ions}\label{App:ZhS-7.1n}

For more precise mathematical model we should take into account the difference between mobilities of the negative and positive ions, \emph{i.\,e.} $\mu_k^{+} \ne \mu_k^{-}$ (see (\ref{ZhSeq-F6})). In particular, if the mobility of ions is different then the values of the function $\psi$, at which the molar charge and molar mobility are equal to zero,  are different. In fact, the molar charge is
\begin{equation}\label{ZhSeq-F34}
   \theta_k(\psi)=\theta_k^{+}(\psi)-\theta_k^{-}(\psi)=\frac{\sinh(\psi-\psi_{k})}{\cosh(\psi-\psi_{k})+\delta_{k}}
\end{equation}
and $\theta_k(\psi_k)=0$.
The molar mobility is
\begin{equation}\label{ZhSeq-F35}
   \mu_k\Theta_k(\psi)=\mu_k^{+}\theta_k^{+}(\psi)-\mu_k^{-}\theta_k^{-}(\psi)=\mu_k\frac{\sinh(\psi-\Psi_{k})}{\cosh(\psi-\psi_{k})+\delta_{k}},
\end{equation}
where
\begin{equation}\label{ZhSeq-F36}
   \mu_k=\sqrt{\mu_k^{+}\mu_k^{-}}, \quad \psi_k-\Psi_k=\frac12 \ln\frac{\mu_k^{+}}{\mu_k^{-}}.
\end{equation}
We call $\Psi_k$ isoionic  point. At $\psi=\Psi_k$ the quantity of the negative and positive ions of substance is coincided.

Difference between $\psi_k$ and $\Psi_k$ is well demonstrated by the example of water ions (see (\ref{ZhSeq-F32})):
\begin{equation}\label{ZhSeq-F37}
   \theta_{_{H_2O}}(\psi)=[\textrm{H}^+]-[\textrm{OH}^-]=2k_w \sinh(\psi-0),
\end{equation}
\begin{equation*}
   2K_w\mu_0\Theta_{_{H_2O}}(\psi)=\mu_{_H}[\textrm{H}^+]-\mu_{_{OH}}[\textrm{OH}^-]=2k_w\mu_0\sinh(\psi-\Psi_0), \quad \Psi_0=\psi_0.
\end{equation*}

In the Tab.~\ref{tab:table6} the mobility of ions $\mu_k^{+}$ and $\mu_k^{-}$ are presented. Data in Tab.~\ref{tab:table6} are taken from software PeakMaster (see \cite{Hirokawa}) that includes a database based on Takeshi Hirokawa's tables with the data of many ions.
\begin{table}[H]
\caption{\label{tab:table6} Parameters of amino acids}

\begin{ruledtabular}
\begin{tabular}{lcccccccccc}
     & $\textrm{pKb}_i$ & $\textrm{pKa}_i$ & $\textrm{pI}_i$ & $\psi_i$ & $\Psi_i$ & $\psi-\Psi_i$& $\Delta\textrm{pI}_i$ & $\delta_i$  & $\mu_i^- $ & $\mu_i^+ $ \\
\hline
Thr  & $2.14$           & $9.200$          & $5.6700$         & $3.062$  & $3.071$  & $-0.009$      & $ 0.004$             & $1694.22 $   & $3.09$     & $3.04$    \\
Pro  & $1.85$           & $10.640$         & $6.2450$         & $1.739$  & $1.633$  & $ 0.106$      & $-0.046$             & $12415.66$   & $2.90$     & $3.58$    \\
Ala  & $2.25$           & $9.857$          & $6.0535$         & $2.179$  & $2.133$  & $ 0.046$      & $-0.020$             & $3180.31 $   & $3.22$     & $3.53$    \\
Iso  & $2.30$           & $9.765$          & $6.0325$         & $2.228$  & $2.183$  & $ 0.045$      & $-0.019$             & $2700.66 $   & $2.67$     & $2.92$    \\
Lei  & $2.26$           & $9.728$          & $5.9940$         & $2.316$  & $2.266$  & $ 0.050$      & $-0.022$             & $2710.00 $   & $2.64$     & $2.92$    \\
Val  & $2.21$           & $9.710$          & $5.9600$         & $2.395$  & $2.353$  & $ 0.042$      & $-0.018$             & $2811.71 $   & $2.84$     & $3.09$    \\
Phe  & $2.13$           & $9.262$          & $5.6960$         & $3.003$  & $2.993$  & $ 0.010$      & $-0.004$             & $1840.64 $   & $2.69$     & $2.74$    \\
Trp  & $2.31$           & $9.594$          & $5.9520$         & $2.413$  & $2.409$  & $ 0.004$      & $-0.002$             & $2192.65 $   & $2.54$     & $2.56$    \\
Met  & $2.13$           & $9.344$          & $5.7370$         & $2.908$  & $2.908$  & $ 0.000$      & $-0.000$             & $2022.88 $   & $2.93$     & $2.93$    \\
Ser  & $2.13$           & $9.302$          & $5.7160$         & $2.957$  & $2.960$  & $-0.003$      & $ 0.001$             & $1927.39 $   & $3.36$     & $3.34$    \\
Gln  & $2.10$           & $9.224$          & $5.6620$         & $3.081$  & $3.067$  & $ 0.014$      & $-0.006$             & $1823.77 $   & $2.88$     & $2.96$    \\
Asn  & $2.10$           & $9.030$          & $5.5650$         & $3.304$  & $3.298$  & $ 0.006$      & $-0.003$             & $1458.71 $   & $3.16$     & $3.20$    \\
$\beta$-Ala
     & $3.42$           & $10.241$         & $6.8305$         & $0.390$  & $0.279$  & $ 0.111$      & $-0.048$             & $1286.68 $   & $3.08$     & $3.85$    \\
Gly  & $2.32$           & $9.780$          & $6.0500$         & $2.188$  & $2.160$  & $ 0.028$      & $-0.012$             & $2685.16 $   & $3.74$     & $3.95$    \\
\end{tabular}
\end{ruledtabular}
\end{table}

\subsection{The Kohlrausch's function}\label{App:ZhS-7.1}

In this section we obtain the analog of the Kohlrausch's function for simplest IEF model (\emph{i.\,e.} at $K_w=0$).
The division of each equation (\ref{ZhSeq-F27}) on the $\mu_k$ and summarization over all $k$ give
\begin{equation}\label{ZhSeq-F38}
   R_t-\varepsilon\Delta S=0,
\end{equation}
where
\begin{equation}\label{ZhSeq-F39}
   R=\sum\limits_{k=1}^{n}\frac{a_k}{\mu_k}, \quad S=\sum\limits_{k=1}^{n}a_k.
\end{equation}
Here $R(x,t)$ is the the analog of the Kohlrausch's function.

Especially interesting is the case, when $\mu_k=\mu$, $k=0,\dots,n$. Then $S=\mu R$ and taking into account the boundary conditions and the initial conditions  (see, for example, \cite{Part1,Part2,Part3}) in 1D case we have problem
\begin{equation}\label{ZhSeq-F40}
   R_t-\varepsilon\mu R_{xx}=0, \quad R_x(0,t)=0, \quad  R_x(L,t)=0, \quad R(x,0)=\frac{1}{\mu}\sum\limits_{k=1}^{n} M_k.
\end{equation}
Obviously, the solution of this problem is
\begin{equation}\label{ZhSeq-F41}
   R(x,t)=\frac{1}{\mu}\sum\limits_{k=1}^{n} M_k=\textrm{const}.
\end{equation}

Notice that near the stationary state, when each concentration is almost localized in its own region, for instance $[x_l,x_r]$ we can write approximation
\begin{equation}\label{ZhSeq-F42}
   \partial_t a_k-\varepsilon\mu_k \partial_{xx} a_k =0, \quad x \in [x_l,x_r].
\end{equation}
This allows to obtain the characteristic time of the steady state release
\begin{equation}\label{ZhSeq-F43}
   t_k \approx  \frac{(x_r-x_l)^2}{\varepsilon\mu_k\pi^2}=\frac{(x_r-x_l)^2 \lambda}{\mu_k\pi^2}.
\end{equation}
For example, at $\lambda=500$, $(x_r-x_l)=0.25$, $\mu_k=1$ we have $t_k\approx 3$.

\subsection{Whether to ignore the Poisson-Boltzmann equation?}\label{App:ZhS-7.2}

In the general case, instead of the electroneutrality equation
\begin{equation}\label{ZhSeq-F44}
   q\equiv\sum\limits \theta_k a_k +2K_w\sinh\psi =0
\end{equation}
and the electric current continuity equation
\begin{equation}\label{ZhSeq-F45}
   \operatorname{div} \boldsymbol{j}=0
\end{equation}
we should use the charge conservation  law
\begin{equation}\label{ZhSeq-F46}
   \partial_t q + \operatorname{div} \boldsymbol{j}=0
\end{equation}
and the Poisson-Boltzmann equation
\begin{equation}\label{ZhSeq-F47}
   \varepsilon_0 \operatorname{div} \boldsymbol{E}=q,
\end{equation}
where $\varepsilon_0$ is the permittivity of water (for water the dimension value of the permittivity is $\varepsilon_0^* \approx 80\cdot 8.854\cdot10^{-12}\, \textrm{F}/\textrm{m}$, $\textrm{F}=\textrm{C}/(\textrm{V}\cdot\textrm{m}$)).

If the permittivity is small enough then (\ref{ZhSeq-F47}) implies (\ref{ZhSeq-F44}) and (\ref{ZhSeq-F46}) implies (\ref{ZhSeq-F45}).
We compare the molar charge of water $([\textrm{H}^{+}]-[\textrm{OH}^{-}])$ and the term $\varepsilon_0 \operatorname{div} \boldsymbol{E}$ in 1D case.

Using the dimensional variables we can write (see, for example, \cite{Part1,Part2,Part3})
\begin{equation*}
 (\varepsilon_0\varphi_{xx})_*=\varphi_{xx} \frac{\varepsilon_0^*E_*}{L_*}=\frac{\varepsilon_0^*R_*T_*}{F_*L*^2}\frac{\varphi_{xx}\lambda}{j_0}\,\,(\textrm{C}/\textrm{m}^3),
 \quad ([\textrm{H}^{+}]-[\textrm{OH}^{-}])_*=2K_w^*F_*\sinh\psi\,\,(\textrm{C}/\textrm{m}^3)
\end{equation*}
or
\begin{equation*}
 (\varepsilon_0\varphi_{xx})_*\approx 2.77\cdot 10^{-8} \varphi_{xx}\frac{\lambda}{j_0}\,\,(\textrm{C}/\textrm{m}^3), \quad
([\textrm{H}^{+}]-[\textrm{OH}^{-}])_*\approx 0.019\sinh\psi\,\,(\textrm{C}/\textrm{m}^3).
\end{equation*}

\begin{figure}[H]
\centering
\includegraphics[scale=0.8]{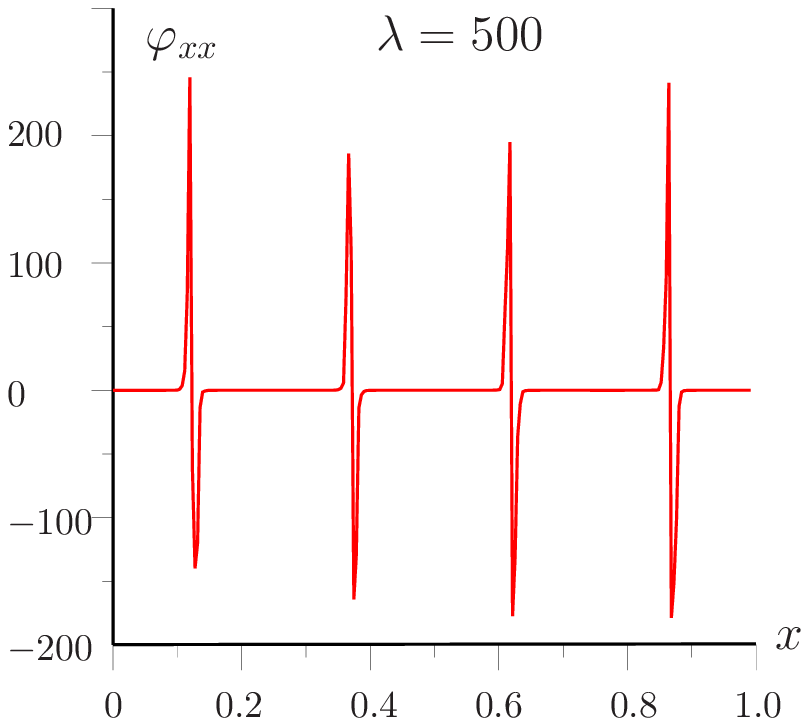}\
\includegraphics[scale=0.8]{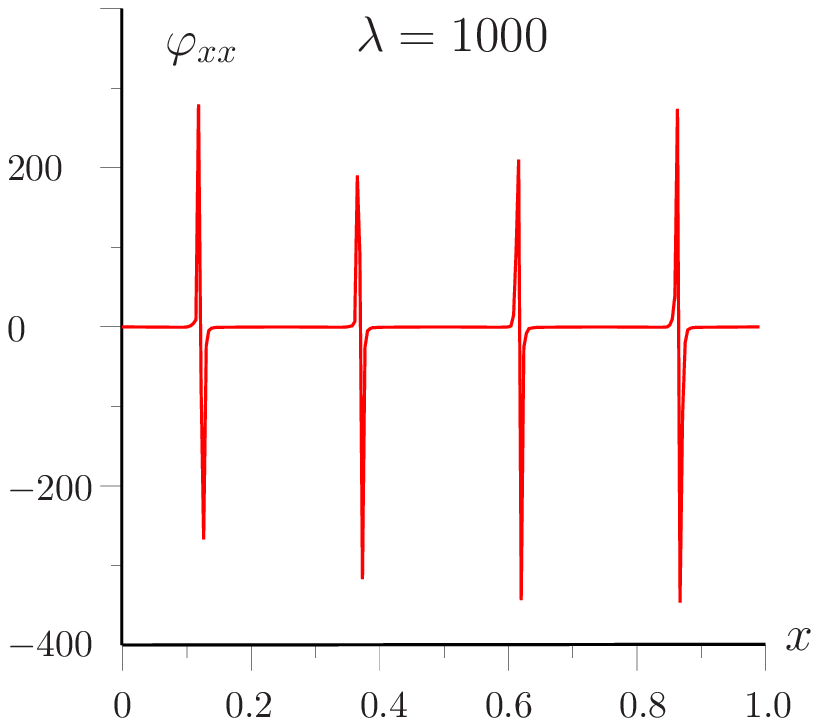}
\caption{The distributions of the $\varphi_{xx}(x)$ at $K_w \ne 0$. The current constant regime ($j=1$) at $\lambda=500$ (left) and $\lambda=1000$ (right),
$t=2.5$, $L_*=2.54\,\textrm{cm}$}
\label{Ris29}
\end{figure}

At $\lambda=500$, $\varphi_{xx} \approx 200$ (see Fig.~\ref{Ris29}, left), and $\psi=1$ we have
\begin{equation*}
 (\varepsilon_0\varphi_{xx})_*\approx 0.0028\,\,(\textrm{C}/\textrm{m}^3), \quad
([\textrm{H}^{+}]-[\textrm{OH}^{-}])_*\approx 0.023\,\,(\textrm{C}/\textrm{m}^3)
\end{equation*}
and at $\lambda=1000$, $\varphi_{xx} \approx 200$ (see Fig.~\ref{Ris29}, right), and $\psi=1$ we have
\begin{equation*}
 (\varepsilon_0\varphi_{xx})_*\approx 0.0056\,\,(\textrm{C}/\textrm{m}^3), \quad
([\textrm{H}^{+}]-[\textrm{OH}^{-}])_*\approx 0.023\,\,(\textrm{C}/\textrm{m}^3).
\end{equation*}
In particular, when $\lambda=1000$ contribution of the term $(\varepsilon_0\varphi_{xx})$ in charge of the mixture only in $4$ times less than the contribution of the water ions.

Thus, if we take into account the water ions, then the using of the Poisson-Boltzmann equation and the charge conservation law instead of the electroneutrality equation
and the electric current continuity equation can play a significant role in the describing of the IEF.

Strictly speaking, the law of charge conservation (\ref{ZhSeq-F46}) is always valid. Indeed, using (\ref{ZhSeq-F5}), similar equation for ions $\textrm{H}^{+}$ and
$\textrm{OH}^{-}$, and (\ref{ZhSeq-F19}) we obtain (\ref{ZhSeq-F46}). If we assume that $q=0$, it is clear that (\ref{ZhSeq-F46}) is splitted into two equations: $q=0$ and $\operatorname{div} \boldsymbol{j}=0$. If we refuse to conditions of electroneutrality mixture, then equation (\ref{ZhSeq-F46}) should be used to define the function $\psi$.

In this case, we rewrite (\ref{ZhSeq-F46}) in the following form
\begin{equation*}
   \sum\limits_{k=1}^{n} \frac{\partial q}{\partial a_k}\,\partial_t a_k + \frac{\partial q}{\partial \psi}\, \partial_t \psi +
   \operatorname{div} \boldsymbol{j}=0.
\end{equation*}
Using (\ref{ZhSeq-F27}), (\ref{ZhSeq-F28}) we get
\begin{equation*}
   \frac{\partial q}{\partial \psi}\, \partial_t \psi +
   \operatorname{div}
\left\{
    \boldsymbol{j}-\sum\limits_{k=1}^{n} \theta_k(\psi)\boldsymbol{i_k}
\right\}=0.
\end{equation*}
Taking into account (\ref{ZhSeq-F30}) and  the relation
\begin{equation*}
\theta'_k(\psi)=\sigma_k(\psi)-\theta^2_k(\psi),
\end{equation*}
finally, we have the evolution equation for the determination of function $\psi$
\begin{equation}\label{ZhSeq-F48}
q_\psi\partial_t \psi +
\operatorname{div} \boldsymbol{J}=r,
\end{equation}
where
\begin{equation}\label{ZhSeq-F49}
q_\psi=
 \sum\limits_{k=1}^{n}  a_k \theta'_k(\psi) +2K_w \cosh\psi,
\end{equation}
\begin{equation*}
\boldsymbol{J}=
\left(
\sum\limits_{k=1}^{n}  \mu_k a_k \theta'_k(\psi) +2K_w \mu_0 \cosh(\psi-\psi_0)
\right)
\left(
\boldsymbol{E}-\varepsilon\nabla\psi
\right),
\end{equation*}
\begin{equation*}
r=-\sum\limits_{k=1}^{n} \boldsymbol{i}_k\cdot\nabla\theta'_k(\psi)=
\sum\limits_{k=1}^{n}
\mu_k \theta'_k(\psi)
\left(
\varepsilon \nabla a_k - \theta_k(\psi)\boldsymbol{E}
\right)\cdot\nabla \psi.
\end{equation*}

Notice that, as expected, the value $\boldsymbol{J}$ formally coincides with the current flux density $\boldsymbol{j}$ for the stationary problem, and the multiplier in front of the term $(\boldsymbol{E}-\varepsilon\nabla\psi)$ is the conductivity $\sigma_{\textrm{stat}}$ for stationary case
(see Sec.~5,  and equations (28) in \cite{Part3} ). Of course,  the contribution of water ions should be added in the conductivity $\sigma_{\textrm{stat}}$.

\section{Conclusion}\label{ZhS-13}

In detail, the described technique of constructing the mathematical models of electrophoresis is presented in \cite{ZhukovBabskiyYudovich,BabZhukYudE}.
In this paper we emphasize the importance of the taking into account the different physical and chemical effects.
Using of the simple models can lead to inadequate description of experiments.

\begin{acknowledgments}
This research is partially supported by Russian Foundation for Basic Research (grants 10-05-00646 and 10-01-00452),
Ministry of Education and Science of the Russian Federation
(programme `Development of the research potential of the high school', contracts  14.A18.21.0873,  8832 and grant 1.5139.2011).

\end{acknowledgments}







\setlength{\bibsep}{4.0pt}

\end{document}